\documentclass[royal]{witpress}
\usepackage{amsmath}
\usepackage{graphicx}
\usepackage{cite}
\title{Techniques for multifractal spectrum estimation in financial time series}
\author{Petr Jizba and Jan Korbel}
\address{Department of Physics, Faculty of Nuclear Sciences and Physical Engineering,
Czech Technical University in Prague, B\v{r}ehov\'{a} 7,\\ 115 19, Prague, Czech Republic}
\begin{document}
\maketitle
\begin{abstract} We show that a multifractal analysis offers a new and potentially promising avenue
for quantifying the complexity of various time series. In particular, we compare the most common techniques used for multifractal scaling exponents estimation. This is done from both theoretical and phenomenological point of view.
In our discussion we specifically focus on methods based on estimation of R\'{e}nyi entropy, which provide a powerful tool especially in presence of heavy-tailed data. As a testbed for the applicability of above multifractal methods we use various real financial datasets,
including both daily and high-frequency data.
\end{abstract}
%

\section{Introduction}
%
During the past few decades, the concepts of scaling and self-similarity have become common fare in various scientific branches, including dynamical systems~\cite{bar}, biological systems~\cite{lovejoy}, quantum field theory~\cite{kleinert} or sociological and economical systems. Global scaling is a key concept, e.g., in theory of critical phenomena and renormalization group, while a global self-similarity is the cornerstone of fractal geometry. Nevertheless, in real systems, such as financial markets, one typically observes local scaling and local self-similarity rules rather than the global ones. A theoretical handle that can successfully deal with such systems is provided by multifractal analysis. This paradigm is based
on the assumption that the distribution of local scaling rules possesses also its scaling rule with characteristic scaling exponent called multifractal spectrum. The theory of multifractals has been deeply studied, e.g., in~\cite{mandelbrot1,mandelbrot2,stanley,harte}. In this connection there have been developed various methods for multifractal spectrum estimation of time series, such as methods based on generalized Hurst exponent~\cite{morales} or wavelet transform~\cite{muzy}. Here we will focus on two most common techniques used for estimation of multifractal scaling exponents, namely Detrended fluctuation analysis~\cite{peng,kandelhardt} and the R\'{e}nyi-entropy-based Diffusion entropy analysis~\cite{scafetta,huang,mfdea}. We compare both methods from the theoretical point of view, and discuss their applicability in time series analysis. To illustrate theoretical results, we apply both methods to the examples of real financial time series, collected on daily and minute basis.

\section{Multifractal analysis}
%
Let us have a time series $\{x_i\}_{i=1}^N$ measured on the specific time lag $s$, (e.g., minute basis or daily basis). We group all points into distinct regions $K_j$, and the probability of occurrence in $j$-th region is given by $p_j = \lim_{N \rightarrow \infty} N_j/N$, where $N_j$ is number of points in $K_j$. We consider that each probability scales with some characteristic exponent, so $p_j \propto s^{\alpha_j}$ and the distribution of scaling exponents is considered in form $\rho(\alpha,s) \propto  s^{-f(\alpha)}$. The scaling exponent $f(\alpha)$ is called \emph{multifractal spectrum} and it is nothing else than fractal dimension of subset with scaling exponent $\alpha$. More details can be found in Refs.~\cite{hentschel,harte}. Alternatively, one can obtain a related representation of characteristic scaling exponents from so-called partition function $Z(q,s) = \sum_j p_j^q \propto s^{\tau(q)}$. Two comments are now in order. Firstly, the relation between multifractal spectrum and scaling exponent of partition function is given by the Legendre transform $\tau(q) = \max_{\alpha} [q \alpha - f(\alpha)]$. Secondly, the exponent $\tau(q)$ is closely related to R\'{e}nyi entropy $S_q(s) = \frac{1}{q-1} \ln \sum_j p_j^q \propto s^{D(q)}$. Eventually, $\tau(q) = \frac{D(q)}{q-1}$. Scaling exponent $D(q)$ is known as a generalized dimension~\cite{hentschel}. R\'{e}nyi entropy was
originally formulated by Hungarian mathematician Alfr\'{e}d R\'{e}nyi~\cite{renyi} in mid 70's and further developed in number of works (see, e.g., Ref.~\cite{jizba} and citations therein).
It represents a one-parametric generalization of Shannon entropy known from information theory and thermodynamics. Because of properties such as observability~\cite{arimitsu} or relation to multifractals, it has found numerous applications in thermodynamics~\cite{jizba}, econophysics~\cite{shefaat} or in quantum mechanics~\cite{dunningham}. In the following section we compare methods based on estimation of moments of probability distribution with the approach based on R\'{e}nyi entropy.

\section{Estimation of multifractal scaling exponents}
%
The first attempt to describe the scaling exponents in time series was done by H.E.~Hurst~\cite{hurst}.
Important upshot of his effort was formulation of the so-called Hurst scaling exponent.
The Hurst exponent is calculated as $\langle |\Delta x(t)| \rangle \propto \Delta t^H$ and it represents a measure of long-memory and persistence. Nowadays, there are many techniques that estimate Hurst exponent, let us just mention  the Rescaled range analysis~\cite{rescaled}. Probably the most popular method called Detrended fluctuation analysis was originally introduced in Refs.~\cite{peng,kandelhardt,hurst2}. The key object of the method is the local \emph{fluctuation function} $f(\nu,s)$, which sums up fluctuations from a local trend. The global fluctuation function  is then calculated as a generalized mean of local fluctuation \mbox{$F(q,s) = \left(\frac{1}{N_s} \sum_\nu^{N_s} f(\nu,s)^q \right)^{1/q}$}. The fluctuation function scales as \mbox{$ F(q,s) \propto s^{h(q)}$}. In some particular cases (e.g. stationarity, positivity), it can be shown that the scaling exponent $h(q)$ is related to conventional multifractal analysis as \mbox{$\tau(q) = q h(q) -1$}.

An alternative approach to estimate the multifractal scaling exponents can be reached through estimation of R\'{e}nyi entropy and its scaling exponent. This method is called Diffusion entropy analysis and it was introduced in Refs.~\cite{scafetta,huang} and further discussed in~\cite{mfdea}. Unlike other methods based on estimation of moments, this approach can successfully deal with distributions with heavy-tails. As an example, let us have a distribution that is self-similar with the scaling form $p(x,t) = \frac{1}{t^\delta} F\left(\frac{x}{t^\delta}\right)$. Then the Shannon entropy ($q \rightarrow 1$) is equal to $S_1(t) = A + \delta \ln t$. In case of multiple scaling exponents, these can be revealed by the whole class of R\'{e}nyi entropies and we obtain $S_q(t) = B_q + \delta(q) \ln t$. At this point, we should stress that proper estimation of R\'{e}nyi entropy for all values of $q$ is crucial in this method. In practice, this boils down to estimation of probability distribution that is constructed via the so-called \emph{Fluctuation collection algorithm}~\cite{scafetta,mfdea}.  To do so, it is necessary to calculate the optimal width of histogram bins so that the interpolating histograms faithfully approximate the underlying distribution. There exist several approaches for such an estimation. We can mention, for instance, the classic rule of Sturges (estimating the optimal number of bins as $1+\log_2 N$), Scott's rule~\cite{scott} or Freedman--Diaconis' rule~\cite{freedman} (both estimate the optimal bin-widths as being proportional to $N^{-1/3}$). Quite recently, a generalization of the above procedure to the case of R\'{e}nyi entropy was presented in Ref.~\cite{mfdea}.
In particular, this provided an estimation of histogram powers $\hat{p}^q$ for several time scales which is necessary in order to compute the corresponding scaling exponents.
\begin{figure}[t]
\centering
\includegraphics[width = 12cm]{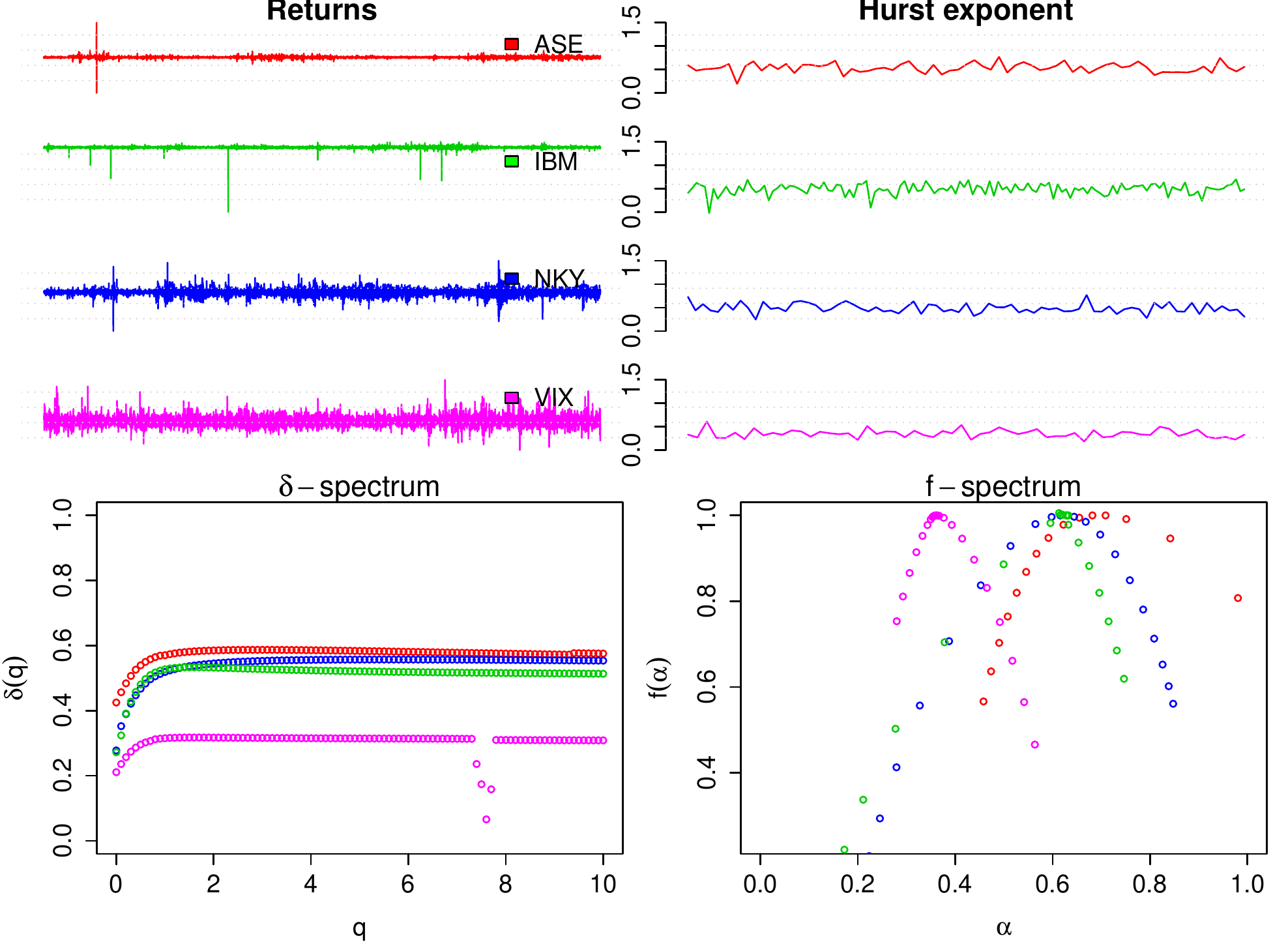}
\caption{Multifractal analysis of daily data. We can observe that the multifractal exponents of the series VIX are different from the other series. This is caused by a volatile nature of the index, resulting also a discontinuity in the $\delta$-spectrum. Note also that the Hurst exponent of VIX series is noticeably lower than in other series.}
\label{fig1}
\end{figure}

By comparing the two methods, we can say that the best approach, when examining data, is to combine both of them and depict both spectra
$f(\alpha)$ and $\delta(q)$. This is because the transformation of the exponents works, strictly speaking, only in the limit $s \rightarrow 0$, which is often replaced by the assumption of exact scaling which is then implemented into the linear regression procedure.  Unfortunately, the exact scaling premiss is not usually fulfilled in case of real time series. Thus, the knowledge of both spectra brings more complex picture of the multifractal nature of the series. Generally we can say that DFA is better, when the estimation of moments is not problematic, while DEA is more fruitful in case of power law distributions and heavy tails. In the next section we compare both spectra and time dependence of the Hurst exponent for various real financial series recorded for both daily and high-frequency ticks.

\section{Applications of multifractal analysis to financial data}
\begin{figure}[t]
\centering
\includegraphics[width = 12cm]{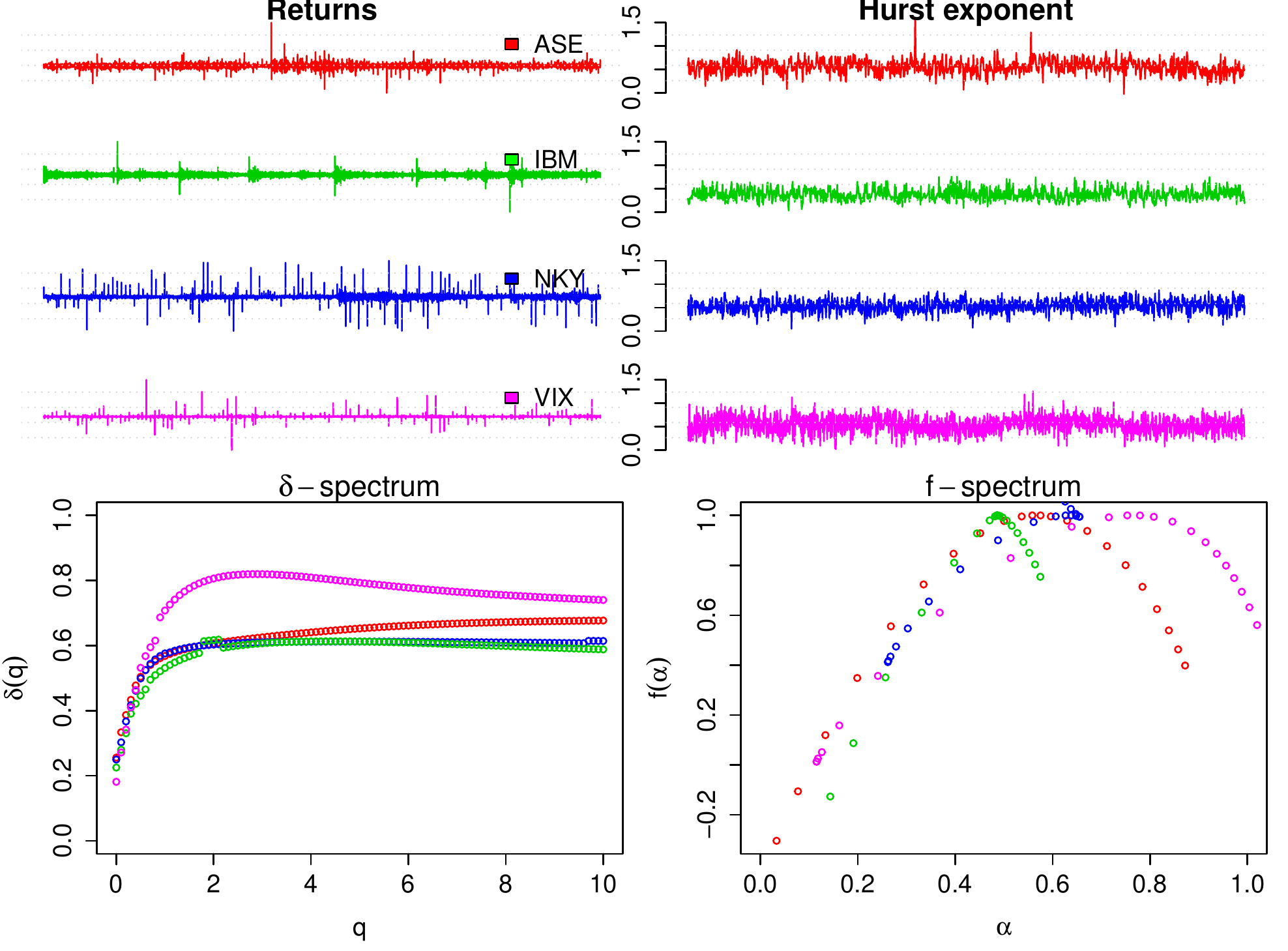}
\caption{Multifractal analysis of high-frequency financial data. Depicted data are highly non-liquid and exhibit a power-law behavior.
This is reflected in discontinuities in both spectra, mainly in the $f$-spectrum of NKY series.}
\label{fig2}
\end{figure}

%
Here we apply the multifractal estimation procedures discussed above to the real financial time series.
Financial markets are for this task an ideal testbed since they represent
open, non-linear and highly structured  complex systems with lots of unexpected and unpredictable phenomena (including sudden jumps, market sentiment, long-memory effects, etc). This in turn brings about a non-trivial multifractal structure of market prices reflected in ensuing time sequences.  We test the applicability of the aforementioned multifractal techniques on several examples of market time series, particularly on stock index Nikkei 225 (index of Tokyo stock exchange), ASE Composite index (main index of Athens stock exchange), IBM stock and VIX index (implied volatility of S\&P 500 options index) on recorded on high-frequency basis and on daily basis. Minute time series were recorded during year 2013 and have approximately $10^5$ tick points, daily data are from the last 10-20 years (depending on the particular series) and have 5000-10000 records. Figs.~\ref{fig1} and~\ref{fig2} depict results of the multifractal analysis of all mentioned series for both characteristic time lags. We can  observe that in case of non-liquid series or anomalous scaling, both methods exhibit certain discontinuities in the spectra. This is caused by the fact that estimation of any quantity based on underlying distribution with heavy tails is usually technically difficult and one needs to make very precise calculations. In case of daily series, we observe that the spectra describe mainly the general characteristic scaling exponents, usually close to $0.5$ (white noise or Wiener process). This can be attributed to the fact that the data are liquid and correlations of the series decay much faster (order of minutes). On the contrary, minute data have much richer structure of scaling exponents, which is apparent mainly from the $\delta$-spectrum. These two figures show that combination of several multifractal methods is desirable and it allows we for a more complete theoretical picture of the observed time series.

\section{Conclusions}
%
The multifractal analysis is an important diagnostic tool that allows to reveal a rich and often intricate scaling structure in variety of complex dynamical systems. In this paper, we have compared several techniques for estimation of multifractal scaling exponents. Notably, we have discussed main theoretical aspects of two currently popular methods, namely Detrended fluctuation analysis and Diffusion entropy analysis and illustrated their utility on the real-life examples of financial time series. The optimal approach is to combine several methods for multifractal spectrum estimation based on different approaches. This can eliminate the numerical and computational artifacts in multifractal spectra that are usually caused by insufficient precision in estimations or by  complex nature of underlying data. Nevertheless, such data are usually the most interesting to investigate. In case of financial series, we have shown that different time-scales have usually different characteristic scaling exponents. This is because the rough coarse-graining can suppress (and often does!) some non-trivial intermediate dynamics alongside with their scaling exponents and hence the effective coarse-grained spectrum is structurally poorer.

\section*{Acknowledgements}
%
Authors want to thank to Xaver Sailer of Nomura Ltd., who provided us with financial data. The work was supported by the Grant Agency
of the CTU in Prague, grant No. SGS13/217/OHK4/3T/14 and the GACR grant No. GA14-07983S.
\bibliographystyle{witpress}
\bibliography{references}
\end{document}